\documentclass{article}
\usepackage{spconf,amsmath,graphicx}
\usepackage{graphicx}
\usepackage{textcomp}
\usepackage{psfrag}
\usepackage{auto-pst-pdf}
\usepackage{xcolor}
\usepackage{enumitem}
\usepackage[absolute]{textpos}
\setlist[itemize]{noitemsep}
\def\BibTeX{{\rm B\kern-.05em{\sc i\kern-.025em b}\kern-.08em
    T\kern-.1667em\lower.7ex\hbox{E}\kern-.125emX}}

\newcommand{\copyrightstatement}{
    \begin{textblock}{15}(0.5,0.7)    
         \noindent
         \centering
         \textblockcolour{white}
         \footnotesize
         \copyright 2023 IEEE. Personal use of this material is permitted. Permission from IEEE must be obtained for all other uses, in any current or future media, including reprinting/republishing this material for advertising or promotional purposes, creating new collective works, for resale or redistribution to servers or lists, or reuse of any copyrighted component of this work in other works
    \end{textblock}}

\title{Processing Energy Modeling\\ for Neural Network Based Image Compression
}

\name{Christian Herglotz, Fabian Brand, Andy Regensky, Felix Rievel, Andr\'e Kaup}
\address{Multimedia Communications and Signal Processing \\
{Friedrich-Alexander-Universit\"at Erlangen-N\"urnberg (FAU)}\\
Erlangen, Germany 
}

\begin{document}
%
\maketitle
\copyrightstatement

\begin{abstract}
Nowadays, the compression performance of neural-network-based image compression algorithms outperforms state-of-the-art compression approaches such as JPEG or HEIC-based image compression. Unfortunately, most neural-network based compression methods are executed on GPUs and consume a high amount of energy during execution. Therefore, this paper performs an in-depth analysis on the energy consumption of state-of-the-art neural-network based compression methods on a GPU and show that the energy consumption of compression networks can be estimated using the image size with mean estimation errors of less than $7\%$. Finally, using a correlation analysis, we find that the number of operations per pixel is the main driving force for energy consumption and deduce that the network layers up to the second downsampling step are consuming most energy. 
\end{abstract}
\section{Introduction}
\label{sec:intro}
In the recent years, image compression using neural networks (NNs) has improved significantly. State-of-the-art compression networks outperform classical methods such as JPEG compression and often show a superior performance to high-end codecs such as HEIC or VVC intra \cite{Xie21,Brand22}. Unfortunately, NNs are very complex due to a large number of parameters and a complex architecture. As a consequence, they are usually executed on highly parallelized GPUs which leads to extremely high processing energies to compress and decompress a picture. As in a recent study it was shown that online video communications contribute significantly to global greenhouse gas emissions \cite{ShiftFull19}, it is clear that this high energy consumption must be mitigated before practical application. 

To this end, we perform a first analysis on the energy consumption of current NN-based image compression solutions, which we call compression networks in the following,  on GPUs. 
Then, we construct an energy model that shows that the energy consumption can be estimated with high accuracy using the resolution of the image. Finally, we compare the compression energy in terms of operations per pixel with various network parameters and find that the complexity in terms of multiply-accumulate (MAC) operations has the highest correlation with the energy consumption. The contributions of this paper are as follows: 
\vspace{-0.3cm}
\begin{itemize}
\item In-depth analysis of the power consumption of image compression networks, 
\item Highly accurate energy modeling approach, 
\item Recommendations for designing energy-efficient networks. 
\end{itemize}
\vspace{-0.3cm}
First, Section \ref{sec:NNs}  introduces the compression networks considered for energy analysis. Section~\ref{sec:meas} explains our energy measurement setup and 
Section \ref{sec:model} shows that the energy consumption can accurately be modeled using a linear model exploiting the number of pixels. Then, Section~\ref{sec:design} analyzes the resulting energy modeling with respect to typical network parameters.  Finally, Section~\ref{sec:conc} concludes this paper.

\section{Neural-Network-based Image Compression}
\label{sec:NNs}
Most current neural network based compression techniques are building on the works by Ball\'e {et al.} \cite{BalleLS2017_Endendoptimized}. The authors proposed to use an autoencoder~\cite{KrizhevskyH2011_Usingverydeep} with an entropy bottleneck. Autoencoders were primarily used for feature extraction, similar to principle component analysis. An autoencoder consists of an analysis transform and a synthesis transform. The former is used to transform the input image to a (typically lower dimensional) latent representation while the latter transforms this representation back to the original image. Since the intermediate latent space is low dimensional, the network is forced to focus this representation on the most important features in order to reconstruct the image as accurately as possible. Both analysis and synthesis transform are typically implemented with multiple strided convolutions. 

Ball\'e {et al.} additionally included an entropy constraint to this latent space. Therefore, a probability distribution for the latent elements is estimated which is then used to estimate the rate required to transmit the latent representation using an asymptotically optimal coder (such as,  e.g., an arithmetic coder). The network is then trained to jointly minimize the distortion and the required rate.

It is easily seen that the quality  of the probability model is crucial for the performance of the compression network. Therefore, subsequent work concerned not only with improving the backbone autoencoder and thereby the feature generation, but also the probability modeling for the latent space. In \cite{BalleMS2018_Variationalimagecompression}, Ball\'e {et. al} proposed a refined probability model, which relies on the so-called hyperprior network, which is a second autoencoder transmitting scale information about the features on a side channel. Thereby, regions with low dynamic information can be compressed more efficiently as the low variance can be taken into account by the probability model. 

This work was further refined by Minnen {et al.}~\cite{MinnenBT2018_JointAutoregressiveHierarchical}, who proposed an autoregressive context model, such that the probability model can take spatial correlation into consideration. The context model consists of causal masked convolutional layers, which together with the hyperprior network estimates mean and variance of a multivariate Gaussian distribution.

The previously mentioned approaches use four strided convolution layers with similar dimensions. In contrast, Cheng {et al.}~\cite{ChengST2020_LearnedImageCompression} proposed an autoencoder structure which is much larger and consists of multiple residual blocks and attention layers. Furthermore, the network 
uses Gaussian mixture models with three Gaussians instead of simple Gaussian models. Thereby, the number of parameters for probability modeling, which need to be estimated, increases.


\begin{table*}[t]
\centering
\caption{Compression networks and main properties. `low l.' and `high l.' represent parameter levels, `padd' is short for `image padding'. The two rightmost columns show the kMAC per pixel (low/high) of the entire and up to the $2^\mathrm{nd}$ layer, respectively.  } 
\resizebox{\textwidth}{!}{ 
\begin{tabular}{l|l|l|l|r|r|r|c||r|r}
\hline
Network name & Abbr.          & Source            & low l. & high l. & low param. & high param. & padd  &  kMAC &  kMAC ($2^{\mathrm{nd}}$) \\ \hline\hline
bmshj2018\_factorized & Bfac &  \cite{BalleLS2017_Endendoptimized} & 1-5        & 6-8         & 2,998,147      & 7,030,531    & $16$   & $73.6$/$163.2$ & $56$/$122.4$\\ \hline
bmshj2018\_hyperprior & Bhyp &  \cite{BalleMS2018_Variationalimagecompression} &1-5        & 6-8         & 5,075,843      & 11,816,323   & $128$  &$76.3$/$169.7$ & $56$/$122.4$\\ \hline
mbt2018\_mean         &  Mmean & \cite{MinnenBT2018_JointAutoregressiveHierarchical}& 1-4        & 5-8         & 7,028,003      & 17,561,699   &$128$  &$80.3$/$181.4$ & $56$/$122.4$ \\ \hline
mbt2018               &  Mcont & \cite{MinnenBT2018_JointAutoregressiveHierarchical}& 1-4        & 5-8         & 14,130,467     & 25,504,596   &$128$  & $166.2$/$201.8$& $122.4$/$122.4$ \\ \hline
cheng2020\_anchor     & Canch & \cite{ChengST2020_LearnedImageCompression} &  1-3        & 4-6         & 11,833,149     & 26,598,956   &$128$  & $373.4$/$834.8$& $345.2$/$771.2$ \\ \hline
cheng2020\_attn       & Cattn & \cite{ChengST2020_LearnedImageCompression}&  1-3        & 4-6         & 13,183,293     & 29,631,788   &$128$   & $415.9$/$930.3$ & $385.1$/$861.1$  \\ \hline
\end{tabular}}
\label{tab:compNets}
\vspace{-.4cm}
\end{table*}
Table~\ref{tab:compNets} summarizes all compression networks evaluated in this work and their main properties. We show proposed quality levels, the number of parameters, and the required operations in kMAC per pixel.
The huge number of parameters hints at the immense complexity of NN-based image compression. 
NNs are thus commonly executed on GPUs leveraging their extreme parallelization capabilities.
This is possible as GPUs provide thousands of compute cores that perform the same instruction on a large number of inputs simultaneously, an operation that is typical for NNs 
~\cite{Owens2008}.
Since Krizhevsky {et al.} first successfully employed GPUs for the training and inference of large-scale neural networks~\cite{Krizhevsky2012}, GPUs have become the de-facto standard for deep learning research and manufacturers have even extended their hardware to provide dedicated tensor cores for these tasks~\cite{NVIDIAVolta, NVIDIATuring}.

\section{Measurement Setup}
\label{sec:meas}
For our measurements, we use desktop PCs (Linux OS) equipped with Nvidia Quadro RTX 4000 GPUs run by Intel Core i5-10505 CPUs. As power meters, we use the internal power meters of the GPU that can be read-out using the NVIDIA SMI interface~\cite{NVIDIA_SMI} at a sampling rate of $f_\mathrm{s} = 10\,$Hz.

\begin{figure}[t]
\psfrag{012}[tc][tc]{ Time [s]}%
\psfrag{013}[bc][bc]{ Power [W]}%
\psfrag{000}[ct][ct]{ $4$}%
\psfrag{001}[ct][ct]{ $5$}%
\psfrag{002}[ct][ct]{ $6$}%
\psfrag{003}[ct][ct]{ $7$}%
\psfrag{004}[ct][ct]{ $8$}%
\psfrag{005}[ct][ct]{ $9$}%
\psfrag{006}[ct][ct]{ $10$}%
\psfrag{007}[bc][bc]{ }
\psfrag{008}[rc][rc]{ $0$}%
\psfrag{009}[rc][rc]{ $50$}%
\psfrag{010}[rc][rc]{ $100$}%
\psfrag{011}[rc][rc]{ $150$}%
\psfrag{GPU Power}[l][l]{GPU Power}%
\psfrag{Measurement Interval12}[l][l]{Measurement Interval}%
\includegraphics[width=0.5\textwidth]{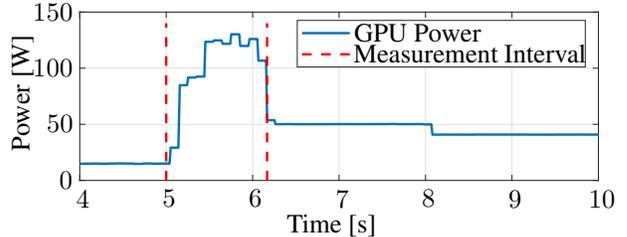}
\vspace{-.9cm}
\caption{GPU power over time (blue) during the execution of the compression network (between the red dashed lines). The network model is loaded to the GPU beforehand.  } 
\label{fig:GPUpow}
\end{figure}

An example for the power consumption of the GPU during the execution of a compression network (model is already loaded to the GPU) is illustrated in Fig.~\ref{fig:GPUpow}. 
The plot shows the power consumption of the GPU as returned by NVIDIA SMI before, during, and after the execution of the compression network. Execution starts at $t=5\,$s and ends at $t=6.2\,$s, as indicated by the red vertical lines.
The higher power value after the execution ($t>6.2\,$s) corresponds to a higher performance state of the GPU~\cite{NVMLvR470}.
As the visible delay in state changes is an inherent property of the power control cycle of the GPU but not part of the compression procedure, it is not taken into account in our measurements.

In our automated measurement script, we perform the following procedure to obtain mean processing energies: First, we perform a reference measurement in which the GPU is in idle mode. For this, we read all power consumption samples over $120$ seconds and calculate the mean power $P_\mathrm{idle}$.

Then, we load the network models on the GPU and preheat the GPU to $80^\circ\,$C by running the current compression network repeatedly until the temperature is reached. Preheating is performed because during the measurement, we execute the compression network multiple times in a row. Preheating ensures that the temperature of the GPU is constant across all executions, such that static power losses are constant, too. 

Afterwards, we start a measurement loop in which the energy consumption $E_\mathrm{comp}$ of compression network execution is measured. For this, we perform $K$ measurements of the same compression modality (compression network, image, and quality level) to ensure statistical validity of the measurement. In each iteration of the $K$ measurements, the modality is executed $M$ times in a row, because the duration of a single execution is often close to the sampling time of the power meter. Consequently, the mean energy is obtained by 
\begin{equation}
\overline{E_\mathrm{comp}} = \frac{1}{K}\sum_{k=1}^K\frac{\big( \sum_{i=1}^N  t_\mathrm{s} \cdot P_{i,k} \big) - E_\mathrm{idle}}
{M}, 
\label{eq:compEnergy}
\end{equation}
where $N$ is the number of power samples from the power measurement interval, $i$ the corresponding sample index, $t_\mathrm{s}= \frac{1}{f_\mathrm{s}}$ the sampling time of the power meter, $P_{i,k}$ the power readout at sample index $i$ from measurement iteration $k$, and $E_\mathrm{idle}$ the corresponding energy consumed in idle mode. 
$E_\mathrm{idle}$ is calculated by $E_\mathrm{idle}=N\cdot t_\mathrm{s}\cdot P_\mathrm{idle}$. 
It is subtracted because we are interested in the dynamic, and not the static energy consumption. 
As measurement interval, we choose the beginning and the ending of the $M$ executions (samples in-between the red vertical lines in Fig.~\ref{fig:GPUpow}). $M$ is chosen such that the total duration of all $M$ executions exceeds $5\,$s. 

The value of $K$ is determined at measurement runtime to ensure statistical validity of the measurement using confidence interval tests as proposed in \cite{Herglotz18}. 
The resulting mean energy $\overline{E_\mathrm{comp}}$ is then used for further analysis and modeling. 

Note that the measurements are performed with the following conditions: First, we test with the GPU in deterministic execution mode to ensure reproducibility~\cite{PyTorch112Repr} which is required in practice. Second, we do not perform entropy encoding and decoding of the latent space, because this is often performed on the CPU. Third, we measure the complete encoder-decoder chain. 
 
We measure the energy for all compression networks listed in Table~\ref{tab:compNets}, where we use the implementations from \cite{compressAI}. For each model, we test all available quality levels on $60$ images from the CLIC image database \cite{CLIC} (test images). Note that the network architectures are provided in two variants with a differing number of parameters: a low level and a high level variant, where the former is used for low quality and the latter for high quality compression. The resolutions of the images span from $751\times500$ to $2048\times1366$. All images are padded according to the restrictions of the compression networks, which means that the height and the width are padded either to multiples of $16$ or $128$. 

To ensure the validity of energy measurements using the internal power meter (NVIDIA SMI), we performed additional experiments using an external power meter, a ZES Zimmer LMG661. We measured the energy consumption of the full desktop PC including the GPU through the main power supply of the PC. In total, we measured the execution of $440$ compression network instances including different networks, images, and deterministic as well as non-deterministic execution. On the power meter, we used the same measurement procedure to derive energies as used for NVIDIA SMI, but we directly read integrated energy values from the power meter as explained in \cite{Herglotz18}. As a result from this benchmark, we find that apart from a constant offset energy, which is related to the PC's peripheral components such as CPU, memory, motherboard etc., NVIDIA SMI's energy measurements are correlated to the power meter's energy measurements with a correlation factor of $0.9986$. We conclude that in the measurement conditions used in this analysis, NVIDIA SMI returns accurate energy measurements.

\section{Energy Modeling}
\label{sec:model}
\begin{figure}[t]
\centering
\psfrag{011}[tc][tc]{ Number of Pixels}%
\psfrag{012}[bc][bc]{ Processing Energy [J]}%
\psfrag{000}[tc][tc]{ $\times10^{6}$}%
\psfrag{001}[ct][ct]{ $0$}%
\psfrag{002}[ct][ct]{ $1$}%
\psfrag{003}[ct][ct]{ $2$}%
\psfrag{004}[ct][ct]{ $3$}%
\psfrag{005}[ct][ct]{ $4$}%
\psfrag{006}[rc][rc]{ $0$}%
\psfrag{007}[rc][rc]{ $20$}%
\psfrag{008}[rc][rc]{ $40$}%
\psfrag{009}[rc][rc]{ $60$}%
\psfrag{010}[rc][rc]{ $80$}%
\psfrag{LQ}[l][l]{\footnotesize{ LQ}}%
\psfrag{HQ}[l][l]{\footnotesize{ HQ}}%
\psfrag{LQ fit}[l][l]{\footnotesize{ LQ fit}}%
\psfrag{HQ fit}[l][l]{\footnotesize{ HQ fit}}%
\includegraphics[width=0.5\textwidth]{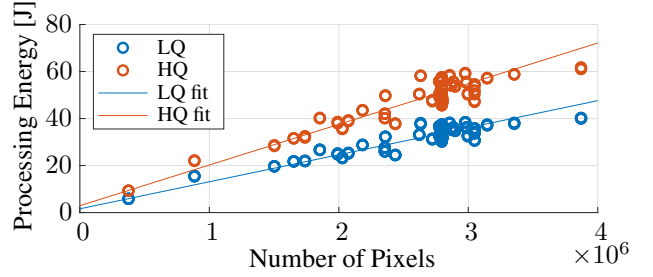}
\vspace{-.8cm}
\caption{GPU energy over number of image pixels for the Bhyp compression network. The red markers correspond to high quality levels (HQ), the blue markers to low quality levels (LQ). The lines are linear least-squares fits to the points.   } 
\label{fig:energy_samples}
\vspace{-.4cm}
\end{figure}

\vspace{-.3cm}
Using the measured data for the six different networks, we attempt to model the energy consumption using high-level properties of the image. To this end, Fig.~\ref{fig:energy_samples} shows the measured energies (NVIDIA SMI) over the number of pixels per image for the Bhyp compression network. 
We can see that if we consider a fixed network architecture including a fixed number of network parameters, e.g., the low quality levels of the Bhyp compression network (blue markers), the measured energy values closely follow a linear relationship with the number of pixels. The same holds true for the high quality levels. The linear correlation yields $0.93$ and $0.92$ for the low and the high quality levels, respectively, with similar results for other compression networks. 

Consequently, we propose to model the energy consumption of neural network processing using a linear equation given by 
\begin{equation}
\hat E = \alpha\cdot S + \beta, 
\label{eq:linModel}
\end{equation}
where $\hat E$ is the estimated energy consumption, $S$ the number of pixels in the image, i.e. the image width multiplied with the image height, $\alpha$ the slope of the linear function, and $\beta$ the offset. Illustratively, $\alpha$ can be interpreted as the mean energy consumption per pixel and $\beta$ as an energy offset per execution. 

To prove the accuracy of the linear estimator from Eq.~\eqref{eq:linModel}, we also calculate the mean relative estimation error of the model \cite{Herglotz18}. It is given by 
\begin{equation}
\bar \varepsilon = \frac{1}{|\mathcal{I}|\cdot |\mathcal{Q}|}\sum_{i\in \mathcal{I}} \sum_{q\in \mathcal{Q}} \frac{|\hat E_{q,i} - E_{q,i}|}{E_{q,i}}, 
\end{equation}
where $i$ is the image index from the complete set of images $\mathcal{I}$ and $q$ the quality level index from the set of quality indices $\mathcal{Q}$, where this set either contains the high qualities or the low qualities. $E_{q,i}$ and $\hat E_{q,i}$ correspond to the measured and the estimated processing energies from Eq.~\eqref{eq:compEnergy} and Eq.~\eqref{eq:linModel}, respectively, of the $i$-th picture and the $q$-th quality level. Training and validation is performed using $10$-fold cross-validation. 

The slope $\alpha$, the offset $\beta$, and the mean estimation errors $\bar \varepsilon$ are summarized for all tested compression networks, separated by high and low quality levels, in Table~\ref{tab:modelVals}. 
\begin{table}[t]
\caption{Modeling parameters and estimation errors for all considered compression networks.  }
\centering
\resizebox{.48\textwidth}{!}{ 
\begin{tabular}{r|r|r|r|r|r|r}
\hline
Network  & Bfac & Bhyp & Mmean & Mcont & Canch & Cattn \\ \hline
$\alpha_\mathrm{low} (\cdot 10^{-5})$ & $1.05$ & $1.15$ & $1.17$ & $1.75$ & $1.99$ & $2.34$\\ 
$\alpha_\mathrm{high}(\cdot 10^{-5})$ & $1.50$ & $1.73$ & $1.76$ & $1.82$ & $4.20$ & $4.97$ \\ \hline
$\beta_\mathrm{low}$ & $1.95$ & $1.61$ & $1.58$ & $3.01$ & $0.37$ & $0.42$ \\ 
$\beta_\mathrm{high}$ & $3.68$ & $2.93$ & $3.17$ & $3.52$ & $2.32$ & $2.12$ \\ \hline
$\bar \varepsilon_\mathrm{low}\,[\% ]$ & $6.09$ & $6.08$ & $6.04$ & $6.30$ & $3.55$ & $3.31$ \\ 
$\bar \varepsilon_\mathrm{high}\,[\% ]$ & $6.11$ & $6.24$ & $6.28$ & $6.46$ & $3.88$ & $3.54$\\ \hline
\end{tabular}}
\label{tab:modelVals}
\vspace{-.4cm}
\end{table}
Considering the estimation error, we can see that the proposed linear model leads to mean estimation errors below $7\%$. For the Canch and Cattn model, the error is significantly lower (below $4\%$). This behavior is probably related to the utilized capacity of the GPU, which is significantly higher for the Canch and Cattn network, as indicated by NVIDIA SMI. This conjecture is an interesting aspect for future research. 

The values of the offset parameter $\beta$ show that there is a significant offset energy needed to execute the compression networks. However, considering image sizes of at least $1120\times 760$ pixels (only a single image from the set has a lower resolution), the offset energy never exceeds a ratio of $20\%$ of the total measured energy. 

The slope parameters $\alpha$ reflect the observation that compression networks for high visual qualities consume more energy than those for low visual quality, because the high-quality values are greater. We can also observe that the highest value is almost five times higher than the lowest value ($4.97$ vs. $1.05$).

\section{Model Analysis}
\label{sec:design}
\vspace{-.3cm}
In the next step, we compare the energy consumption (in energy per pixel) with network parameters to identify correlations, which could be helpful in network design decisions. Therefore, we analyze the per-pixel-complexity of the networks in the unit multiply-accumulate (MAC) operations per pixel. We calculate the MAC for all networks by counting the operations in each convolutional layer and weighting them according to the current downsampling at this layer. We report the values in Table~\ref{tab:compNets}. 

\begin{figure}
\psfrag{013}[tc][tc]{ Multiply-Accumulate operations in kMAC}%
\psfrag{014}[bc][tc]{ Slopes $\alpha$}%
\psfrag{000}[tc][tc]{}
\psfrag{001}[ct][ct]{ $0$}%
\psfrag{002}[ct][ct]{ $200$}%
\psfrag{003}[ct][ct]{ $400$}%
\psfrag{004}[ct][ct]{ $600$}%
\psfrag{005}[ct][ct]{ $800$}%
\psfrag{006}[ct][ct]{ $1000$}%
\psfrag{007}[rc][rc]{ $0$}%
\psfrag{008}[rc][rc]{ $1$}%
\psfrag{009}[rc][rc]{ $2$}%
\psfrag{010}[rc][rc]{ $3$}%
\psfrag{011}[rc][rc]{ $4$}%
\psfrag{012}[rc][rc]{ $5$}%
\psfrag{Bfacaaaa}[l][l]{ Bfac}%
\psfrag{Bhyp}[l][l]{ Bhyp}%
\psfrag{Mauto}[l][l]{Mmean}%
\psfrag{Mcont}[l][l]{ Mcont}%
\psfrag{C}[l][l]{ Canch}%
\psfrag{Cattn}[l][l]{ Cattn}%
\psfrag{low}[l][l]{$\alpha_\mathrm{low}$}%
\psfrag{high}[l][l]{$\alpha_\mathrm{high}$}%
\psfrag{interp}[l][l]{ Lin. fit}%
\includegraphics[width=0.5\textwidth]{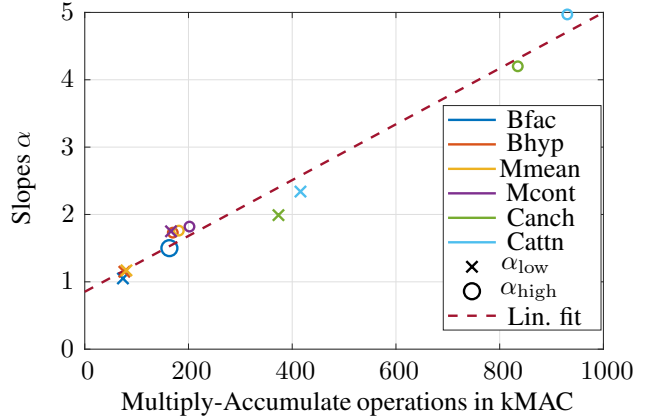}
\vspace{-0.8cm}
\caption{Slopes $\alpha$ over the MAC for all compression networks (color) and their quality levels (x'es for low qualities, o's for high qualities). The dashed line is a linear fit for all points.  } 
\label{fig:slopeOverMac}
\end{figure}

To show that the MAC is an important parameter in the energy consumption, we plot the MAC values and the $\alpha$ values for all networks in Fig.~\ref{fig:slopeOverMac}. 
We can see that all parameters closely follow a linear relationship with the complexity of the network in terms of MAC. Next to a linear term, we can also find an offset, which indicates that a minimum energy per pixel is required (e.g., due to memory read and write of the raw image). 

We further study the kMAC per network layer to find reasons for the high power consumption of an image compression network. In particular, we focus on the part of the network that is most important for processing complexity. Here, we find that the layers up to and including the second downsampling step are very important (see the corresponding kMAC values in the last column of Table~\ref{tab:compNets}). Since the deeper layers are performed on rapidly decreasing resolution, the impact on the kMAC per pixel is small. Performing a similar fit to these $2^\mathrm{nd}$-layer MAC values as to the entire MAC values, we find that the linear correlation is almost as high (MRE of $7.64\%$ for the entire MAC vs. $11.16\%$ for the $2^\mathrm{nd}$-layer MAC). We conclude that these layers are the major reason for a high energy consumption of neural-network based image coders, such that they should be the main target when developing energy efficient NN-based compression schemes.

\section{Conclusions}
\label{sec:conc}
\vspace{-.3cm}
In this paper, we investigated the energy consumption of state-of-the-art neural-network based image compression networks on a GPU.
We showed that the energy consumption linearly depends on the number of pixels and that the network layers up to the second downsampling step contribute most to the overall energy consumption. 

In future work, we plan to analyze further networks, to separately analyze the encoding and the decoding process, and to additionally consider entropy encoding and decoding. Finally, we will test other GPUs and optimized hardware to generate a fair comparison with legacy image compression methods.




%


\bibliographystyle{IEEEbib}
\bibliography{literature}
%

\end{document}